# Thermodynamical insight on the role of additives in shifting the equilibrium between white and grey tin

Nikolay Dementev

*Department of Chemistry, Temple University, Philadelphia, PA 19122*

**Abstract**

In this study methods of statistical thermodynamics were applied to white tin – grey tin equilibrium to find if there is any correlation between Debye temperature of an additive and its effect on the equilibrium. Gibbs free energy of the transition was calculated for a composite system comprised of tin and the additive in assumption of Debye temperature of the composite system being a linear combination of Debye temperatures of the components with their respective molar fractions as weighing coefficients.

Calculations showed that additives which Debye temperatures are above $214^{o}K$ should stabilize grey tin, while the additives with Debye temperatures below $214^{o}K$ should stabilize the white one. Devised rule corroborates with all previously reported experimental data on the effect of different additives on the transition. Predictions of the rule about non reported additives provide a mean for its experimental verification.

1. **Introduction**

It is well known that at temperatures below 13.2°C, β-Sn (white tin) transforms into α-Sn (grey tin).[1,2] Abrupt increase of the tin's specific volume (~26%) during the transition usually leads to disintegration of a specimen, an effect also known as a tin plague or a tin pest.[1-3] Potential danger a tin pest might present in the constructions held by the tin based alloys, inspired an extensive search for additives that might inhibit the disadvantageous transition.[4-10] It has been found that lead (Pb) is the best inhibitor amongst the known ones.[4] However, environmental and health considerations call to halt use of Pb, thus urging one to search for its replacement.[4] Studies in this area, however, were mostly empirical and based on a try-and-see principle so far, as no leading rule in searching for the additives was devised yet.[4-10]

Results of consideration of the effect of additive on a white to grey tin transition with respect to the additive's Debye temperature are reported here. Gibbs free energy associated with the transformation was calculated using the methods of statistical thermodynamics. It was found that additives with Debye temperatures above and below $214^\circ K$ should stabilize grey and white tin, respectively.

2. **Outline of calculations**

Firstly, in order to check if the calculation path is correct, methods of the statistical thermodynamics will be applied to the transition without any additives:

$$Sn_\beta \to Sn_\alpha \qquad (1)$$

Secondly, the temperature of the zero change of Gibbs free energy will be calculated for a composite system with an additive:

$$(Sn_\beta + \text{additive}) \to (Sn_\alpha + \text{additive}) \qquad (2)$$

Calculations will be conducted in a similar manner as for an additive free system, assuming that the Debye temperature of each of the components of the composite system can be presented as a composite property constructed as a linear combination of Debye temperature of pure component and the Debye temperature of an additive (Fig.1):

$$\Theta(Sn_\beta + \text{additive}) \equiv \Theta_\beta^* = x(\Theta_A - \Theta_\beta) + \Theta_\beta \quad (3)$$

$$\Theta(Sn_\alpha + \text{additive}) \equiv \Theta_\alpha^* = x(\Theta_A - \Theta_\alpha) + \Theta_\alpha \quad (4)$$

$$x = \frac{\#\text{of additive atoms}}{\#\text{of additive atoms} + \#\text{of tin atoms}} \quad (5)$$

where $\Theta_\beta$ and $\Theta_\alpha$ are Debye temperatures of a pure white tin and a grey tin, respectively;

$\Theta_A$ is Debye temperature of a pure additive;

$\Theta_\alpha^*, \Theta_\beta^*$ are *effective* Debye temperatures of (grey tin + additive) and (white tin + additive) systems, respectively;

$x$ is a molar fraction of an additive in composite systems.

Finally, some major conclusions about the correlation between Debye temperature of the additive and its effect on the transition will be made.

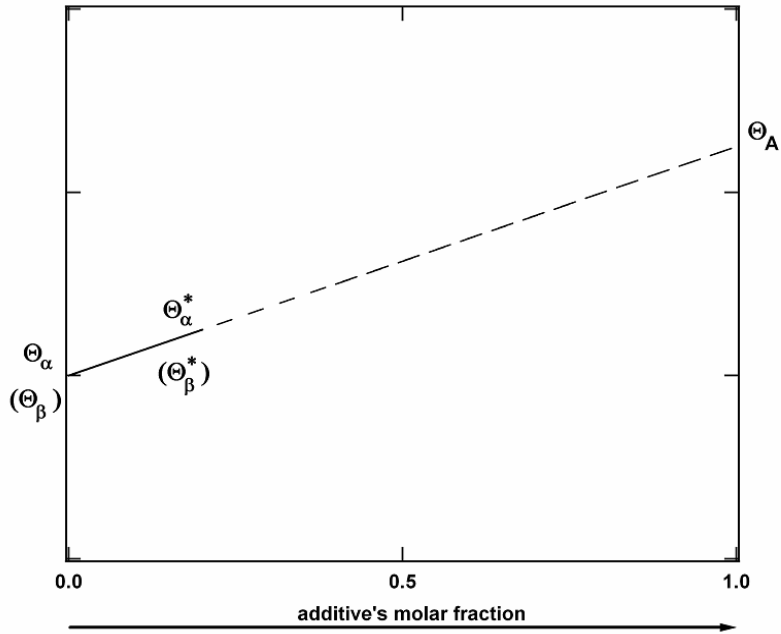

**Figure 1.** Dependence of the effective Debye temperature on additive's content

3. **Transition without additives**

Change of the Gibbs free energy of the system during the heating at constant pressure for each of the modifications can be expressed as:

$$G^T - G^\circ = -kT \ln Q \quad (6)$$

where  $G^T$  is Gibbs free energy at temperature $T$ (°K);

$G^\circ$  is Gibbs free energy at 0 K;

$k$  is Boltzmann's constant;

$Q$  is a canonical partition function

$T$  is an absolute temperature

Thus, the change of the Gibbs free energy for grey tin will be:

$$G_\alpha^T - G_\alpha^\circ = -kT \ln Q_\alpha \quad (7)$$

And for the white tin:

$$G_\beta^T - G_\beta^\circ = -kT \ln Q_\beta \quad (8)$$

And, finally, for the phase transition $Sn_\beta \to Sn_\alpha$:

$$\Delta G^T = G_\alpha^T - G_\beta^T = -kT \ln\left(\frac{Q_\alpha}{Q_\beta}\right) + G_\alpha^\circ - G_\beta^\circ \quad (9)$$

Canonical partition function, in general, can be expressed as a product of the molecular partition functions:

$$Q = q_{transl} \, q_{rot} \, q_{vib} \, q_{elec} \quad (10)$$

where $q_{transl}$, $q_{rot}$, $q_{vib}$, and $q_{elec}$ are the translational, rotational, vibrational, and electronic partition functions, correspondingly.

Since the phase transition $Sn_{white} \to Sn_{gray}$ occurs in solid state at relatively moderate temperature, the only contribution from the vibrational mode $q_{vib}$ can be considered.

Thus, one can write down for diatomic molecule:

$$Q \cong q_{vib} = \frac{e^{(-h\nu/2kT)}}{1-e^{(-h\nu/kT)}} \tag{11}$$

where  $h$  is Planck's constant;

$\nu$  is the frequency of oscillations

Eq. (11) can be rewritten as:

$$Q \cong q_{vib} = \frac{e^{(-\Theta/2T)}}{1-e^{(-\Theta/T)}} \tag{12}$$

where  $\Theta = \frac{h\nu}{k}$  is Debye temperature of the substance

There are *three* vibrational modes for every atom in the crystal. That is why partition function for 1 mole of atoms in the crystal can be presented as:

$$Q = q_{vib}^{3N_a} \tag{13}$$

where  $N_a$  is Avogadro's number

Thus, eq. (9) can be rewritten as:

$$\Delta G_m^T = G_{m(\alpha)}^T - G_{m(\beta)}^T = -3RT \ln\left(\frac{q_{vib(\alpha)}}{q_{vib(\beta)}}\right) + G_{m(\alpha)}^\circ - G_{m(\beta)}^\circ \tag{14}$$

$$\Downarrow$$

$$\Delta G_m^T = -3RT \ln\left[\frac{1-e^{(-\Theta_\beta/T)}}{1-e^{(-\Theta_\alpha/T)}} e^{\left(\frac{\Theta_\beta - \Theta_\alpha}{2T}\right)}\right] + G_{m(\alpha)}^\circ - G_{m(\beta)}^\circ \tag{15}$$

where  $R$  is a gas constant (8.31 J/mol·K);

subscript $m$  denotes *molar* functions;

$\Theta_\beta$  is a Debye temperature of a white tin (200°K);[3,11]

$\Theta_\alpha$  is a Debye temperature of a grey tin (230°K) [3,11]

In order to do calculations based on eq.(15), it is necessary to find out value of the $G_{m(\alpha)}^\circ - G_{m(\beta)}^\circ$ term. This term can be derived from eq.10 and the fact that equilibrium temperature for the white tin-grey tin system is 13.2°C (286.35°K) [3]:

$$\Delta G_m^{286.35} = -3 \times 286.35 R \ln\left[\frac{1-e^{(-\Theta_\beta/286.35)}}{1-e^{(-\Theta_\alpha/286.35)}} e^{\left(\frac{\Theta_\beta-\Theta_\alpha}{2\times286.35}\right)}\right] + G_{m(\alpha)}^\circ - G_{m(\beta)}^\circ = 0 \quad (16)$$

$$\Downarrow$$

$$G_{m(\alpha)}^\circ - G_{m(\beta)}^\circ = 3 \times 286.35 R \ln\left[\frac{1-e^{(-\Theta_\beta/286.35)}}{1-e^{(-\Theta_\alpha/286.35)}} e^{\left(\frac{\Theta_\beta-\Theta_\alpha}{2\times286.35}\right)}\right] \quad (17)$$

$$G_{m(\alpha)}^\circ - G_{m(\beta)}^\circ = -1044.08 \ (J/mol) \quad (18)$$

Eq.(15) can be rewritten as:

$$\Delta G_m^T = -3RT \ln\left[\frac{1-e^{(-\Theta_\beta/T)}}{1-e^{(-\Theta_\alpha/T)}} e^{\left(\frac{\Theta_\beta-\Theta_\alpha}{2T}\right)}\right] - 1044.08 \quad (19)$$

The function $\Delta G_m^T(T)$, calculated from eq.(19), is shown in Figure 2. As one can see (Fig.2) the dependence has a positive slope, meaning that grey tin is more stable than the white one at temperatures below 286.35°K ($\Delta G_m^T < 0$), and white tin is more stable than the grey one at temperatures above 286.35°K ($\Delta G_m^T > 0$). These results are in full agreement with the experimental observations.[1,2]

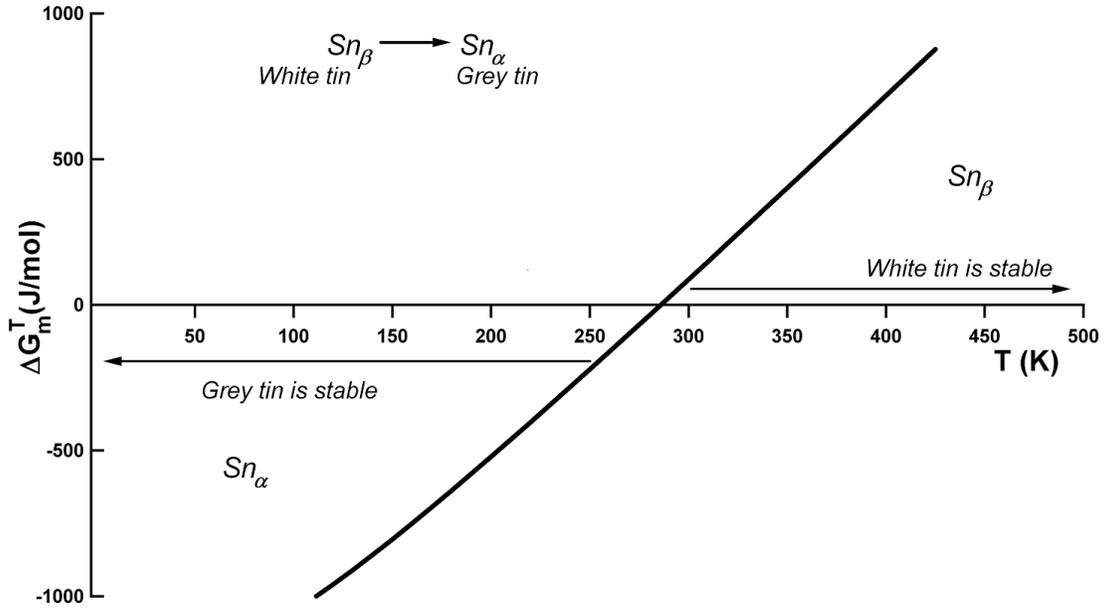

**Figure 2.** Dependence of free Gibbs energy of the transition $Sn_\beta \rightarrow Sn_\alpha$ on temperature, calculated from eq.(19)

## 4. Transition with an additive

In order to describe transition $(Sn_\beta + \text{additive}) \rightarrow (Sn_\alpha + \text{additive})$, Eq.(19) should be transformed into the following one:

$$\Delta G_m^{286.35} = -3 \times 286.35 R \ln\left[\frac{1-e^{(-\Theta_\beta^*/286.35)}}{1-e^{(-\Theta_\alpha^*/286.35)}} e^{\left(\frac{\Theta_\beta^* - \Theta_\alpha^*}{2 \times 286.35}\right)}\right] - 1044.08(1-x) \quad (20)$$

Eq.(20) allows one to determine the change of the free Gibbs energy of the transition $(Sn_\beta + \text{additive}) \rightarrow (Sn_\alpha + \text{additive})$ at 286.35K as a function of molar fraction $x$ and the Debye temperature $\Theta_A$ of the additive (see an outline of calculations).

Results of the calculations of $\Delta G_m^{286.35}$ as a function of $\Theta_A$ for $x = 0.01; 0.025; 0.05; 0.1 \text{ and } 0.2$ are presented in the Fig.3.

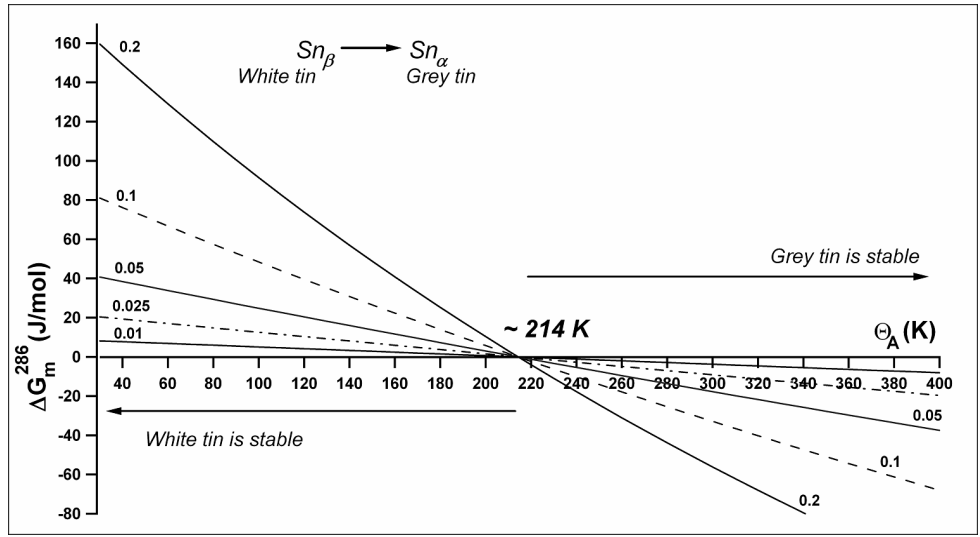

**Figure 3.** Dependence of the change of free Gibbs energy of the transition $(Sn_\beta + \text{additive}) \to (Sn_\alpha + \text{additive})$ on Debye temperature of an additive for its different molar fractions (0.2, 0.1, 0.05, 0.025, 0.1) at 286.35K

One can notice that all curves plotted in Fig.3 have negative slopes and intersect $\Theta_A$ axis at ~214° K, meaning that all additives with $\Theta_A < 214°K$ should stabilize a white tin ($\Delta G_m^{286.35} > 0$), and all additives with $\Theta_A > 214°K$ should stabilize a grey tin ($\Delta G_m^{286.35} < 0$).

5. **Verifications and predictions**

Debye temperatures of different additives with respect to 214K are presented in Fig.4.[11] As one can see, the predictions of the rule, derived above, are in full agreement with all of the experimental results found in literature (Figure 4, solid lines). It was found that Pb, Bi, Sb, and Cd stabilize white tin [1,4,8], while Al, Ge, Cu, Zn, and Ag stabilize grey tin. [4,8]

The rule also allows one to suggest that additives: Tl, In, Yb, β–La, Th, Au, Gd, U and Lu (Figure 4, dashed lines), which effect on $Sn_\beta \to Sn_\alpha$ transition was not investigated yet (to the best of the author's knowledge), will inhibit $Sn_\beta \to Sn_\alpha$ transition.

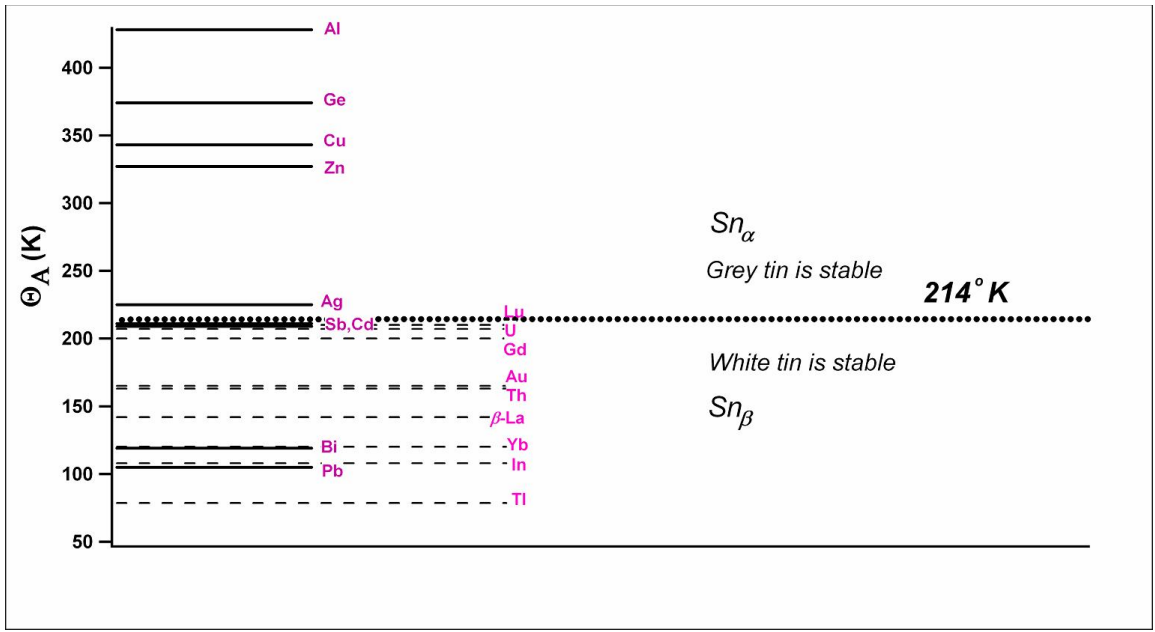

**Figure 4.** Debye temperatures of additives, which effect was (solid lines) and was not (dashed lines) experimentally verified with respect to 214K (dotted line)